# Spin transport parameters of NbN thin films characterised by spin pumping experiments

K. Rogdakis[1,*], A. Sud[1], M. Amado[2], C. M. Lee[2], L. McKenzie-Sell[2], K.R. Jeon[2], M. Cubukcu[1], M. G. Blamire[2], J. W. A. Robinson[2], L. F. Cohen[3], and H. Kurebayashi[1,†]

[1]London Centre for Nanotechnology, University College London, London WC1H 0AH, United Kingdom

[2]Department of Materials Science & Metallurgy, University of Cambridge, Cambridge CB3 0FS, United Kingdom

[3]The Blackett Laboratory, Imperial College London, London SW7 2AZ, United Kingdom

**Abstract**

We present measurements of ferromagnetic-resonance - driven spin pumping and inverse spin-Hall effect in NbN/$Y_3Fe_5O_{12}$ (YIG) bilayers. A clear enhancement of the (effective) Gilbert damping constant of the thin-film YIG was observed due to the presence of the NbN spin sink. By varying the NbN thickness and employing spin-diffusion theory, we have estimated the room temperature values of the spin diffusion length and the spin Hall angle in NbN to be 14 nm and $-1.1 \times 10^{-2}$, respectively. Furthermore, we have determined the spin-mixing conductance of the NbN/YIG interface to be 10 $nm^{-2}$. The experimental quantification of these spin transport parameters is an important step towards the development of superconducting spintronic devices involving NbN thin films.

**Introduction**

The extraction of key functional materials parameters associated with electron transport is important for the development of new solid-state device schemes as well as testing prototypes. In the field of spintronics, the spin Hall angle ($\theta_{SH}$) represents the strength of spin-Hall effect (SHE) [1] that converts charge currents into spin currents via the relativistic spin-orbit interaction. The spin diffusion length ($l_{SD}$) [2] is a parameter that describes the distance over which non-equilibrium spin currents can diffuse before dissipation and is crucial in determining the useful device dimensions of future spintronic applications. Moreover, the spin angular momentum transfer across a ferromagnetic (FM) and non-magnetic (NM) interface can be parameterised by the spin mixing conductance ($g_r^{\uparrow\downarrow}$) which governs the spin current generation efficiency in spin pumping processes [3]. These spin transport parameters can be determined by employing different measurement techniques. For example, it is possible to use lateral spin-valves to quantify $l_{SD}$ and $\theta_{SH}$ in non-magnetic materials [4, 5, 6, 7]. Spin pumping [3, 8, 9] is another established method to investigate spin transport parameters in a variety of materials, such as metals [10], inorganic [11, 12] and organic semiconductors [13, 14], graphene [15] and topological insulators [16]. It should be noted that spin pumping relies on the transfer of angular momentum from a ferromagnet with precessing moments into an adjacent non-magnetic layer, and does not suffer from the conductance mismatch problem which causes difficulties in electrical spin injection through ohmic contacts [11]. Using a FM conductor as spin injector in a spin pumping experiment can potentially give rise to microwave (MW)-induced photo-voltages [17] due to time-varying resistance changes produced by the magnetic precession coupled with a time-varying current, as well as the ISHE in the FM layers [18, 19]. The use of FM insulators such as $Y_3Fe_5O_{12}$ (YIG) to conduct spin pumping experiments has the advantage because these effects are negated. In addition, YIG has a low bulk Gilbert damping constant ($\alpha \simeq 6.7 \times 10^{-5}$) and a high Curie temperature ($T_C = 560$ K), enabling efficient spin pumping at room temperature (RT) [20].

In this paper, we report spin pumping in thin-film YIG/NbN bilayers with the aim of extracting multiple spin transport parameters of NbN thin films in the normal state. NbN is a key material for superconducting (SC) spintronics [21] with a bulk $T_C$ of approximately 16.5 K, a SC energy gap of 2.5 meV, and a SC coherence length of 5 nm [22]. NbN is increasingly used in the field of SC spintronics, for example in spin-filter Josephson junctions [23, 24, 25] and to demonstrate spectroscopic evidence for odd frequency (spin-triplet) superconductivity at the interface with GdN [26]. Recently, Wakamura *et al.* observed an unprecedented enhancement of the SHE at 2K, interpreted in terms of quasiparticle mediated transport [27]. Quasiparticle spin transport has also been investigated by spin pumping and by monitoring the spin Seebeck effect [28, 29]. To the best of our knowledge, spin transport parameters in NbN such as $l_{SD}$ and $\theta_{SH}$ have only been extracted by Wakamura *et al.* [27] by the spin absorption method in lateral spin-valves, and it is vitally important to extract these parameters also by other characterisation techniques and with NbN grown by different growth methods. This can, for example, help to understand whether spin transport parameters in NbN have a significant dependence on the growth conditions. In our study, by using high-quality epitaxial thin-film YIG it is possible to observe a modulation of the Gilbert damping constant ($\alpha$) with NbN thickness and therefore extract $l_{SD}$ of NbN (14 nm) and $g_r^{\uparrow\downarrow}$ of the YIG/NbN interface (10 nm$^{-2}$). Furthermore, we have investigated the NbN-thickness-dependence of the ISHE voltage ($V_{ISHE}$) and have determined $\theta_{SH}$ of NbN (-1.1 ×10$^{-2}$) by the spin pumping technique. We compare $l_{sd}$ extracted by three independent methods, namely the thickness dependence of α and $V_{ISHE}$ as well as Hanle spin precession, and we find good agreement between them. Determining the normal-state spin-transport parameters in NbN from spin-pumping-induced ISHE is important, which enables the comparison between parameters extracted using various techniques from different research groups [e.g. 27-29]. By accumulation of a body of results, we will then be able to understand the fundamental nature of SHE and spin transport in NbN which can be useful and transferable to future spintronics research using SC NbN [21, 30].

**Material growth**

Epitaxial YIG thin films are grown on (111)-oriented GGG single crystal substrates by pulse laser deposition (PLD) in an ultra-high vacuum chamber (UHV) with a base pressure better than $5\times10^{-7}$ mbar. Prior to film growth, the GGG substrate are ultrasonically cleaned by acetone and isopropyl alcohol and annealed ex-situ at 1000 ºC in a constant $O_2$ flow environment for 8 hours. The YIG is deposited from a stoichiometric (polycrystalline) target using a KrF excimer laser (248 nm wavelength), with a nominal energy of 450 mJ and fluence of 2.2 W cm$^{-2}$ in 0.12 mbar of $O_2$ at 680 ºC, and pulse frequency of 4 Hz for 60 minutes. The YIG is post-annealed at 750 ºC for 1.5 hours in 0.5 mbar partial pressure of static $O_2$ and subsequently cooled to RT at a rate of -10 K/min. Atomic force microscopy (AFM) reveals that a root-mean-squared roughness of the YIG films is less than 0.16 nm over 10×10 µm scan size [Fig. 1(a)]. The YIG films were characterised by a SC quantum interference device (SQUID) magnetometer and have a saturation magnetisation ($M_S$) of 140 ± 3 emu cm$^{-3}$ [Fig. 1(b)], which matches the bulk value [31]. In Fig. 1(c) we have plotted a high-angle X-ray diffraction trace of the same film where Laue fringes indicate layer-by-layer growth of YIG and good lattice-matching with the substrate. Figure 1(d) shows low-angle X-ray reflectivity from a YIG film and from the decay and angle separation of the Kiessig fringes, we determined a nominal thickness $t_{YIG} = 60 \pm 2$ nm. Following the growth of YIG, films were directly transferred in air to a UHV sputter deposition system with a base pressure of $1\times10^{-9}$ mbar. NbN is grown by reactive sputtering in a gas mixture of argon (72%) and nitrogen (28%) with the deposition rate of 85 nm min$^{-1}$. The growth temperature is RT, giving polycrystalline NbN layers. We grew NbN with different thicknesses ($t_{NbN}$) from 5 to 50 nm.

**Ferromagnetic resonance (FMR) setup and spin pumping measurements**

FMR is performed using a broadband coplanar waveguide (CPW) and ac-field modulation technique as illustrated in Fig. 2(a). The samples are placed face down on top of the CPWs where an insulator tape is used for electrical insulation. We generate dc ($H$) and ac ($h_{ac}$) magnetic fields by electromagnets and the absorbed power at the modulation frequency is measured by a MW power detector and a lock-in amplifier while $H$ is swept. An input MW power ($P_{MW}$) of 100 mW is used unless otherwise is stated. We kept the modulation field amplitude ($h_{ac}$) smaller than the measured FMR linewidths of all samples tested, in order to avoid artefacts by strong modulation. The magnetic field is applied along different in-plane and out-of-plane directions related to the samples as shown in Fig. 2(a). The FMR absorption ($V_P$) was measured using a MW power detector for different frequencies typically ranging from 2-12GHz as depicted in Fig. 2(b) (for a sample with $t_{NbN}$ = 10 nm). For each scan, the resonance field ($H_{res}$) and the half-width-at-half-maximum linewidth ($\Delta H$) of the FMR signal are determined by a fit using differential forms of symmetric and anti-symmetric Lorentzian functions (Appendix A). Figure 2(c) shows the frequency dependence of the extracted $H_{res}$ for different NbN thicknesses. The curves of the frequency dependence for all samples, including $t_{NbN}$ = 0 nm, overlap suggesting no significant modification of the YIG magnetic anisotropy due to the presence of NbN. We note here that the effective magnetisation ($M_{eff}$) extracted from the fits for each sample shows larger values than the $M_s$ value measured in the SQUID. This enhanced $M_{eff}$ has often been observed in other thin-film studies [32, 33] and a detailed understanding of this lies outside of the scope of the present work. For spin transport analysis discussed later, we use the values extracted by SQUID measurements since it is a more direct measurement of magnetisation, while we confirmed that the discrepancy between $M_s$ and $M_{eff}$ does not alter the calculated spin transport parameters significantly. Although the magnetic anisotropies of the YIG films are unchanged with or without the presence of NbN, the magnetization relaxation of YIG represented by $\Delta H$ shows a clear dependence on $t_{NbN}$ as shown in Fig. 3(a). With a linear fit to the data for each thickness using $\Delta H = \Delta H_0 + (4\pi\alpha/\gamma)f$ where $\Delta H_0$ and $\gamma$ describe the inhomogeneous broadening and the gyromagnetic ratio respectively, we have quantified α for each sample as shown in Fig. 3(b). $\alpha = (5.4 \pm 0.2) \times 10^{-4}$ was obtained for bare YIG, which compares well to previously reported values [34, 35]. A gradual increase of α is observed with increasing NbN thickness, in agreement with spin pumping through the YIG/NbN interface where the α dependence with $t_{NbN}$ is given by [36]:

$$\alpha(t_{NbN}) = \alpha_0 + \left(\frac{g_L \mu_B g_r^{\uparrow\downarrow}}{4\pi M_s t_{YIG}}\right) \cdot \left[1 + \frac{g_r^{\uparrow\downarrow} \rho_{NbN} l_{sd} e^2}{2\pi \hbar \tanh\left(\frac{t_{NbN}}{l_{sd}}\right)}\right]^{-1} \qquad (1).$$

Here, $\alpha_0$ is the Gilbert damping constant for $t_{NbN}= 0$ nm and the second term represents the damping enhancement by spin pumping into NbN; $g_L$ is the free electron Landé factor which is assumed equal to 2, $g_r^{\uparrow\downarrow}$ is the effective real-part spin-mixing conductance across the NbN/YIG interface; $\rho_{NbN}$ is the resistivity of NbN which was measured for each sample [see inset of Fig. 3(b)], and $e$ is the electron charge. A best fit of the data in Fig. 3(b) using Eq. (1) yields $g_r^{\uparrow\downarrow} =$ 10 ±2 nm$^{-2}$ and $l_{sd}$ =14 ± 3 nm. The extracted $l_{sd}$ can be well compared with the value (7 nm) by Wakamura *et al.* [27] using the spin-absorption method in lateral spin-valve devices. We also found that the spin coupling of NbN/YIG is as good as heavy-metals/YIG interfaces since $g_r^{\uparrow\downarrow}$ is comparable to those of YIG/Pt, YIG/Ta and YIG/W [35]. We note from analytic calculations (Appendix B) that the additional damping expected from eddy currents cannot explain the observed NbN thickness dependence of α.

We now discuss the ISHE voltage ($V_{ISHE}$) measurements. In Figs. 4(a) and 4(b) we show typical data sets for $V_{ISHE}$ (for direct comparison we present also corresponding $V_p$ data) for $t_{NbN}$ = 20 nm and $f$ = 3 GHz. Note that, since we used the lock-in ac field-modulation method for both detections, the curves represent the derivative of the signals without the ac field-modulation: for both $V_P$ and $V_{ISHE}$ a symmetric Lorentzian lineshape is expected without the ac field modulation. As expected from spin pumping and ISHE, we observe a clear $V_{ISHE}$ peak at the YIG precession frequency. By changing the sign of $H$ [observe the sign of magnetic field axis for Figs. 4(a) and 4(b)], we observe a sign change of $V_{ISHE}$ in agreement with the symmetry of spin pumping [37]. Corresponding measurements for $t_{NbN}$ = 5 nm are depicted in Figs. 4(c) and 4(d). By using the known ac field modulation amplitude as well as differential forms of symmetric and anti-symmetric Lorentzian functions (Appendix A), we quantify the peak amplitude of ISHE voltage defined as $V_{ISHE}^*$. The $P_{MW}$-dependence of $V_{ISHE}^*$ shown in Figs. 5(a) and 5(b) suggests that $V_{ISHE}^*$ is proportional to $P_{MW}$, consistent with standard spin-pumping theory [36].

We have also performed $H$ - angular dependent measurements of $V^*_{ISHE}$ along in-plane and out-of-plane directions of the NbN/YIG films. The in-plane angular dependence of the spin pumping experiment follows the expression $V_{ISHE}^* \propto \boldsymbol{\epsilon_x} \cdot (\boldsymbol{J_s} \times \boldsymbol{\sigma}) \cdot |\boldsymbol{\sigma} \times \boldsymbol{h_{rf}}|$ where the first part is due to the ISHE symmetry, $E_{ISHE} \propto (\boldsymbol{J_s} \times \boldsymbol{\sigma})$, multiplied by the amplitude of magnetic torque generated by MW-induced magnetic field $|\boldsymbol{\sigma} \times \boldsymbol{h_{rf}}|$; here, $\boldsymbol{\epsilon_x}$ is the unit vector along x direction in the measurement's framework shown in Fig. 2(a). The first component gives a $\cos\theta$ dependence whereas the second produces a $|\cos\theta|$ dependence, which combined nicely matches our experimental results shown in Fig. 6(a). The rationale to plot $V_{ISHE}^* t_{NbN}/\rho_{NbN}$ against $t_{NbN}$ is to include the thickness dependence of $\rho_{NbN}$ allowing to fit the data points based on bare NbN as well as those of the YIG/NbN bi-layers. In addition, this analysis can display the asymptotic behaviour of the data/fit-curves towards the long thickness limit. The in-plane symmetry re-confirms that spin rectification effects are not a dominant mechanism in our measurements since in this case a higher order $\sin 2\theta$ component is expected in the voltage symmetry [17]. We also measured the out-of-plane angular dependence of $V_{ISHE}^*$ as shown in Fig. 6(b) and moreover we applied the Hanle precession model [38] to fit our data. In this case the out-of-plane $V_{ISHE}^*$ is given by:

$$V_{ISHE}^*(\phi) \propto \left\{\cos(\phi)\cdot\cos(\phi-\phi_M) + \sin\phi\cdot\sin(\phi-\phi_M)\cdot\left[\frac{1}{1+(\omega_L\cdot\tau_s)^2}\right]\right\} \quad (2)$$

$\omega_L = g_L\mu_B \cdot (\mu_0 H)/\hbar$ is the Larmor frequency and $\tau_s$ is the spin relaxation time in NbN; $\phi$ and $\phi_M$ represent the angle of between the z-axis and $H$ and the equilibrium magnetic moment direction, respectively. By minimizing the total magnetic energy of the FM layer consisting of the Zeeman and demagnetization energy, the following equation is derived to determine the value of $\phi_M$ with respect to $\phi$: $\phi_M = \phi - \arctan\left[\text{sgn}(\phi)\cdot\sqrt{\left(\frac{\cos(2\phi)+\left(\frac{\mu_0 H_{res}}{\mu_0 M_{eff}}\right)}{\sin(2\phi)}\right)^2 + 1} - \left(\frac{\cos(2\phi)+\left(\frac{\mu_0 H_{res}}{\mu_0 M_{eff}}\right)}{\sin(2\phi)}\right)\right]$ [39]. After spin currents are injected inside NbN, they start precessing due to the externally applied $H$. This is described by the well-known Hanle precession model which is the basis of Eq. (2). The equilibrium spin orientation depends on the precession rate ($\omega_L$) and the spin relaxation rate ($1/\tau_s$), both of which contribute in the equation. When $\tau_s$ is much shorter than $1/\omega_L$, the injected spins do not precess and instead generate $V_{ISHE}$ with spin orientation along $M$ ($\phi_M$). This is the case for the red curve in Fig. 6(b). In the opposite extreme condition (depicted as blue curve in Fig. 6(b)), spins precess many times and dephase along the $H$ orientation ($\phi$), resulting in an approximately $\cos(\phi)$ angle dependence. Fitting the data in Fig. 6(b) using Eq. (2) allows us to estimate $\tau_s$. In particular, the best fit of the measured $V_{ISHE}^*(\phi)$ was obtained giving an extracted $\tau_s = 11 \pm 2$ ps. This value quantified by the Hanle model can be compared with $\tau_s$ independently calculated from the spin diffusion model as already discussed above, i.e. $\tau_s = (l_{sd})^2/D$ where $D$ is the Einstein diffusion coefficient (its value equal to 0.4−0.56 cm$^2$/s was taken from Ref. [40]). Following this approach and by using $l_{sd}$=14 nm as extracted from the thickness dependence of damping modulation, we calculated $\tau_s = 3.6\text{-}5.9$ ps which is a fair agreement between the two different $\tau_s$ extraction methods.

In the following section, the $\theta_{SH}$ of NbN is determined from the thickness dependence of $V^*_{ISHE}$ as shown in Fig. 7. Using the spin transport parameters discussed above and Eq. (3), we estimate the spin current emitted at the NbN/YIG interface, $j_s$, as well as the value of $\theta_{SH}$ extracted by fitting the thickness dependence of $V^*_{ISHE}$ [39]:

$$V^*_{ISHE} = \left(\frac{w_y \rho_{NbN}}{t_{NbN}}\right) \cdot \theta_{SH} l_{sd} \cdot \tanh\left(\frac{t_{NbN}}{2l_{sd}}\right) \cdot j_s \quad (3)$$

$$\text{where } j_s = \left(\frac{G_r^{\uparrow\downarrow}\hbar}{8\pi}\right) \cdot \left(\frac{\mu_0 h_{rf} \gamma}{\alpha}\right)^2 \cdot \left[\frac{\mu_0 M_s \gamma + \sqrt{(\mu_0 M_s \gamma)^2 + 16(\pi f)^2}}{(\mu_0 M_s \gamma)^2 + 16(\pi f)^2}\right] \cdot \left(\frac{2e}{\hbar}\right)$$

$$\text{with } G_r^{\uparrow\downarrow} \equiv g_r^{\uparrow\downarrow} \cdot \left[1 + \frac{g_r^{\uparrow\downarrow} \rho_{NbN} l_{sd} e^2}{2\pi\hbar \tanh\left(\frac{t_{NbN}}{l_{sd}}\right)}\right]^{-1}.$$

Here we assume that YIG is a perfect insulator; $\mu_0 h_{rf}$ is the amplitude of MW magnetic field (56 µT for 100 mW); $w_y$ is defined by the width of MW transmission line. For the data fitting procedure we use $\theta_{SH}$ and $l_{sd}$ as free parameters, where the best fitting was achieved for 1.1 ×10$^{-2}$ and 14 nm, respectively. We also confirmed the sign of $\theta_{SH}$ to be negative by comparing YIG/NbN data with a YIG/Pt control sample where Pt is known to have a positive $\theta_{SH}$ [1]. We emphasise that the value of $l_{sd}$ extracted by the thickness dependence of $V^*_{ISHE}$ agrees very well with the one extracted from the thickness dependence of damping. The former approach includes spin-orbit and spin-transport properties of NbN, whereas the latter is purely related with magnetic properties of YIG. We found that the value we extract by our spin pumping experiments is similar to $\theta_{SH}$ quantified by Wakamura *et al.* using lateral spin-valve samples ($\theta_{SH}$ ~-1× 10$^{-2}$) [27] for the temperature region between 20 to 200 K. Although there is difference in temperature between experiments by Wakamura *et al.* and ours, an agreement of the same sign and magnitude in $\theta_{SH}$ quantified by different techniques (*i.e.* spin pumping and spin-absorption) has been observed. The value of $\theta_{SH}$ of the same material but grown and measured by different research groups can vary rather significantly, for example as in the cases of Pt [41] and some topological insulators [42, 43, 44]. Such differences might result from variation in sample quality where the density of scattering impurities can particularly influence $\theta_{SH}$ via the extrinsic spin-Hall mechanisms [1]. We note that the resistivity of NbN used in the Wakamura *et al.* study measured at 20 K (220 µΩcm) is roughly three times greater than our NbN films at the same temperature (65 µΩcm). This highlights that the resistivity and mobility of NbN might be highly growth-dependent, possibly due to the stoichiometry of Nb and N as well as the nitrogen vacancy concentration. The NbN spin-Hall resistivity of Wakamura *et al.* is 2.2 µΩ·cm at 20 K [27], whereas our spin-Hall resistivity at RT is calculated to be 0.5 µΩ·cm which is smaller owing to the resistivity difference. For the relevance of SC spintronics, we also compare our $\theta_{SH}$ value with those of Nb thin films reported in previous works. Morota *et al.* measured $\theta_{SH}$ of several 4d and 5d transition metals by the spin absorption method in the lateral spin valve structures [6] including Nb. They quantified $\theta_{SH}$ of Nb to be -8.7 ×10$^{-3}$ at 10K, which is close to our $\theta_{SH}$ in NbN at RT. There is recent work by Jeon *et al.* who measured $\theta_{SH}$ = -1×10$^{-3}$ in Nb at RT [39]. Direct comparison between $\theta_{SH}$ of Nb and NbN is not possible but they are within the same order, suggesting that there are similar atomistic spin-orbit contributions from Nb atoms both for Nb and NbN. Details of this will be further clarified when more realistic theoretical studies of SHE in NbN become available.

As a final remark, we also performed FMR measurements as a function of temperature to determining the low-temperature spin-pumping properties of NbN through the SC $T_c$. However, a significant increase of magnetic damping was observed as the temperature was lowered (this behaviour is summarised in Appendix C). This enhanced damping complicated the investigation of $V_{ISHE}$ across the SC $T_c$.

**Conclusions**

We determined the spin transport parameters of polycrystalline NbN thin-films by the spin pumping technique using epitaxial YIG thin-films at RT. We observe a modification of the YIG Gilbert damping parameter as a function of the variation of the NbN film thickness, confirming spin current injection in the NbN layer. By applying a spin-diffusion model, we have estimated $l_{sd}$ =14 ± 3 nm in NbN and $g_r^{\uparrow\downarrow}$ = 10 ±2 nm$^{-2}$ at the NbN/YIG interface. From the NbN thickness dependence of the ISHE voltages, we determine $\theta_{SH}$ to be equal to -1.1 ×10$^{-2}$. We also compare $l_{sd}$ of NbN extracted by three different techniques (thickness dependence of both α and $V_{ISHE}$ as well as the Hanle measurements) and found good agreement between them. The measured parameters are a good reference to understand the NbN spin-orbit and spin transport properties and to aid the design of feasible spintronic experiments/devices in the normal and SC state.

**Acknowledgment** This work was supported by the Engineering and Physical Sciences Research Council through the Programme Grant "Superspin" (Grant No. EP/N017242/1) and International Network Grant (Grant No. EP/P026311/1).

**Appendix A: Derivation of FMR fit curves**

In normal dc FMR analysis, the measured dc voltage can be decomposed into symmetric and anti-symmetric Lorentzian functions with respect to $\mu_0 H_{res}$, with weights of $A_{sym}$ and $A_{asy}$ respectively, where combined lead to the following general power absorption expression [which is applicable both for FMR absorption ($V_p$) and ISHE voltage ($V_{ISHE}$)]:

$$P_{dc}(H) = A_{sym}(H) + A_{asy}(H) + V_0 = A_{sym}\frac{\Delta H^2}{(H-H_{res})^2+\Delta H^2} + A_{asy}\frac{\Delta H(H-H_{res})}{(H-H_{res})^2+\Delta H^2} + V_0, \quad (4)$$

where $V_0$ is a background voltage. The first term gives the symmetric lineshape and the second term produces the anti-symmetric one. For FMR measurements based on ac magnetic-field modulation, where an additional pair of coils on electromagnets provide small ac magnetic field, $P_{ac}$ has the following relationship with $P_{dc}$.

$$P_{ac} = \frac{dP_{dc}}{dH}h_{ac} \quad (5)$$

where, $h_{ac}$ is the amplitude of ac magnetic field modulation. With these two equations, we can calculate $P_{ac}$ as:

$$P_{ac}(H) = -A_{sym}h_{ac}\frac{2(H-H_{res})\Delta H^2}{\{(H-H_{res})^2+\Delta H^2\}^2} - A_{asy}h_{ac}\frac{\Delta H\{(H-H_{res})^2+\Delta H^2\}}{\{(H-H_{res})^2+\Delta H^2\}^2} \quad (6)$$

This equation was used to fit the ac field modulated signals, both $V_p$ and $V_{ISHE}$, in our study. The first term gives now the anti-symmetric lineshape and the second term produces the distorted symmetric one. Figure 8 (a) and (b) display typical FMR data together with best fit curves using Eq. (4) and (6), respectively, with corresponding extracted parameters presented in Fig 8 as legends. We also checked that there was no experimental artifact by doing our ac experiments, by directly confirming that ac (Fig. 8a) and dc (Fig. 8b) measurements for the same experimental conditions generate the same fit parameters.

**Appendix B: A simplified model for the eddy-current damping**

We consider a slab of magnet containing a chain of distributed magnetic moments *m* as shown in Fig 9(a). In order to model the eddy-current damping in NbN, we first calculate the magnetic flux at point P where the distance between the point and the slab is *x* (Fig 9a). We can estimate the magnetic field at point P generated by a moment at (0, y) using the Biot-Savart law, as:

$$B = \frac{\mu_0}{4\pi}\frac{m}{(x^2+y^2)^{3/2}} \quad (7)$$

where $\mu_0$ is the permeability of free space. We assume that the length of the chain is infinitely long, which is a valid assumption by taking in consideration that the film thickness is much shorter than the sample lateral dimensions. By integrating the contribution of the individual moments, we calculate the total magnetic field $B_{total}$ as:

$$B_{total} = 2\int_0^\infty \frac{\mu_0}{4\pi}\frac{m}{(x^2+y^2)^{3/2}}dy = \frac{\mu_0}{2\pi}\frac{m}{x^2} \tag{8}$$

Using this $B_{total}$ expression within this quasi-2D picture, we can calculate the magnetic flux $\Phi$ at point P. By definition, $\Phi = \iint B_{total}ds$, where the integration surface is defined by the thickness $t_{NbN}$ and the width $w$ of the NbN film. This reads:

$$\Phi = \iint B_{total}ds = w \times \int_{t_{YIG}/2}^{\frac{t_{YIG}}{2}+t_{NbN}} \frac{\mu_0}{2\pi}\frac{m}{x^2}dx = \frac{\mu_0 w m}{\pi}\frac{t_{NbN}}{t_{YIG}(t_{YIG}+2t_{NbN})} \tag{9}$$

For the definition of the integration region, we assume that the chain of the magnetic moments is located at the centre of the YIG film.

After estimating the magnetic flux, we can calculate the radiative dissipation power $P$ as:

$$P = \frac{\omega}{2Z_{NbN}}\Phi^2 = \frac{\omega}{2Z_{NbN}}\left(\frac{\mu_0 w m}{\pi}\frac{t_{NbN}}{t_{YIG}(t_{YIG}+2t_{NbN})}\right)^2 \tag{10}$$

Here $Z_{NbN}$ is the impedance of the NbN film and for simplification we assume that the real part (resistance) dominates, meaning that $Z_{NbN} \approx R_{NbN} = \rho_{NbN}(d/wt_{NbN})$. Using the total non-equilibrium magnon energy generated during the experiments as $\hbar\omega NV$ (here, $N$ is the number of the non-equilibrium magnons and V is the volume of YIG), we can express the rate of energy dissipation being:

$$\frac{1}{\tau} = \frac{P}{E} = \frac{\omega\, wt_{NbN}}{2\rho_{NbN}d\hbar\omega N}\left(\frac{\mu_0 w m}{\pi}\frac{t_{NbN}}{t_{YIG}(t_{YIG}+2t_{NbN})}\right)^2 \tag{11}$$

Finally, the damping component caused by eddy currents generated by the time-dependent flux change can be given by:

$$\alpha_{eddy} = \frac{1}{2\omega}(1/\tau) = \frac{wt_{NbN}}{4\rho_{NbN}d\hbar\omega NV}\left(\frac{\mu_0 w m}{\pi}\frac{t_{NbN}}{t_{YIG}(t_{YIG}+2t_{NbN})}\right)^2 \tag{12}$$

As this model is a simplified one, we only discuss $\alpha_{eddy}$ qualitatively. In particular, we can extract the NbN thickness dependence of $\alpha_{eddy}$ by using this expression and find that it is proportional to $\left(\frac{t_{NbN}^{3/2}}{t_{YIG}+2t_{NbN}}\right)^2$. We plot the dependence in Fig. 9 (b) which indicates that the damping based on this mechanism should monotonically increase with thickness. However, this trend is different from what we experimentally observed, where $\alpha$ becomes constant for the larger thickness limit. This suggest that the damping mechanism through the eddy current in the NbN layers is not significant and can be neglected for the examined NbN thicknesses. Moreover, in the work by Flovik et al. [45] they discuss the eddy current effect on the lineshape of the FMR spectrum. They showed that when eddy currents exist in an FM/NM bi-layer, the FMR lineshape can be significantly affected, varying from a pure symmetric shape to a mixture of symmetric and anti-symmetric ones. Experimentally, we have not observed strong $A_{asy}$ component, suggesting that the eddy current in our NbN film does not play a significant role in our measurements. In addition, similar eddy current and radiative damping mechanisms has also been discussed by Schoen et al. [46]. They demonstrated that when their sample is placed 100 μm away from the waveguide, radiative damping with the waveguide is largely supressed. Since we also inserted an insulating tape between our samples and the waveguide, we believe that the radiative damping is minor in our experiments. Furthermore, Qaid et al. [47] reported that although eddy-current damping can be observed in a weak spin-orbit material (in their case a conducting polymer), this is not the case for a high spin orbit metal (Pt). For instance, they showed that the damping enhancement in a YIG/Pt structure can still be dominated by the spin-pumping effect in Pt. Since our NbN is a sufficiently high spin-orbit material, we believe that the eddy-current component is much smaller (an order of magnitude at least) than that of spin-pumping into NbN.

**Appendix C: Low temperature measurements of spin pumping in NbN/YIG samples**

It is widely reported that YIG thin-films tend to show significant temperature dependent magnetic damping [32, 33, 48, 49], where the superb damping character at RT is lost when the films are cooled to lower temperatures. The origin of this remains under debate but enhanced low temperature two-magnon scattering (due to interfacial defects in ultrathin films) [32] in combination with rare-earth or $Fe^{2+}$ impurity scattering [50, 51] are likely mechanisms. Jermain *et al.* [33] discuss that, if the FMR linewidth has a peaked temperature-dependence that dominates over the proportionality expected with $M_S(T)$ increase, impurity scattering is the more likely mechanism. Although the nature of the impurities remains ambiguous, other reports of the high frequency characterisation of PLD-grown and sputtered YIG thin films have pointed out the likely significance of $Gd^{3+}$ diffusion from the GGG substrate [52, 53, 54].

Our own extensive FMR measurements of bare YIG on GGG (of comparable thicknesses) [55] show that, when $Gd^{3+}$ impurities are concentrated in a thin (1-5 nm) layer near the substrate interface, they form a ferromagnetic sublattice that, as its moment increases at low temperatures, opposes the net YIG magnetisation [50, 56], and also introduces magnetic disorder and additional damping channels that dominate the film's FMR response.

Here we describe the low-temperature characterisations of our YIG/NbN samples. Figure 10 summarises both FMR absorption spectra and ISHE voltages as a function of temperature for the sample with NbN thickness of 10 nm. With decreasing temperature, there is a clear increase of $\Delta H$, leading to a corresponding reduction of the FMR absorption signal, as shown in Fig. 10(a). The FMR spectrum at 3K can be extracted by taking multiple scans to improve the signal to noise ratio through data averaging. Figure 10(b) shows that $\Delta H$ increases by a factor of 5 between 300K and 3K, with a steep enhancement below 100 K. For direct comparison we present data in Fig. 10(b) both of YIG/NbN (black points) and bare YIG samples (red points). It is clear that linewidth enhancement at low temperatures is due to YIG. In comparison with the previous low temperature FMR studies on YIG, we can detect an FMR signal down to 3K in the MW transmission geometry, possibly owing to a relatively thick film. Unfortunately, the $\Delta H$ enhancement significantly hindered our ISHE detection plotted in Figs. 10(c) and (d). The voltage peak is comparable or below the noise level at 50 K and it was not possible to investigate the evolution of $V_{ISHE}$ across the $T_c$ of NbN which is 11 K for the 10 nm film, measured by the four point dc resistance $R^{4p}$ shown in the inset of Fig.10(d). To study the spin transport properties in NbN across $T_c$, by spin-pumping technique, would require an improvement of YIG thin-film quality to overcome the observed $\Delta H$ enhancement. We note that a very recent work by Umeda *et al.* exploited the spin-Seebeck effect as a spin-injection method, demonstrating an interesting coherent peak in spin-Seebeck coefficient related to the quasi-particle spin transport [29].

[*]rogdakis7@gmail.com

[†]h.kurebayashi@ucl.ac.uk

**Figure captions**

FIG.1: Structural and magnetic properties of a bare (111)-oriented YIG film (nominally 60-nm-thick) used in this work and deposited onto GGG. (a) 10×10 µm$^2$ AFM topography scan showing a root-mean-square roughness of less than 0.16 nm. (b) Magnetization hysteresis loops characterised by a superconducting quantum interference device magnetometer showing a saturation volume magnetization of $140 \pm 3$ emu cm$^{-3}$. (c) High angle X-ray diffraction data demonstrating (111) orientation with visible Laue fringes on the (444) and (888) diffraction peaks characteristic of layer-by-layer growth. (d) Low angle X-ray reflectometry data (black) with a best-fit (red) curve from which we estimate a nominal thickness of $60 \pm 2$ nm.

FIG.2: (a) A schematic of the spin-pumping setup. The lateral area of all samples is 5x5mm$^2$. MW magnetic fields ($h_{rf}$) were generated by the transmission line to generate magnetic dynamics in the YIG film. Spin currents ($j_s$) were emitted at the YIG/NbN interface, which can induce ISHE voltages detected through the two electrodes attached to the edges of the sample. We simultaneously measured the FMR absorption signal as a voltage in a microwave power meter ($V_P$) connected to the microwave line and the ISHE signal ($V_{ISHE}$) using two lock-in amplifiers. (b) FMR absorption measurements for different MW frequencies. (b) FMR absorption measurements for different MW frequencies. Voltages in our MW power detector were measured while magnetic fields were swept. Dots in red, green, blue, cyan, pink, yellow and black represents measurement results for 3, 4, 5, 6, 8, 10 and 12 GHz respectively. (c) A plot of frequency versus FMR field ($H_{res}$) for samples with different NbN thicknesses. Dots represent experimental results and curves are produced by fitting using the Kittel formula.

FIG.3: (a) Frequency dependence of FMR linewidth of YIG/NbN samples with different NbN thicknesses. Experimental data (filled points) is fitted by a linear line $\Delta H = \Delta H_0 + (4\pi\alpha/\gamma)f$,

where $\Delta H_0$ and $\gamma$ describe the inhomogeneous broadening and the gyromagnetic ratio respectively, from which the Gilbert damping coefficient, $\alpha$, is extracted. (b) Plots of $\alpha$ for different YIG/NbN samples. Equation (1) was used to fit to the thickness dependence with the spin-diffusion length and the real part of mixing conductance as fitting parameters. The inset depicts the resistivity as a function of NbN thickness.

FIG.4: ISHE measurements. Simultaneous measurements of FMR absorption and ISHE voltages for positive (a) and negative (b) magnetic field values for a $t_{NBN} = 20$ nm sample. Corresponding data for $t_{NBN} = 5$ nm are depicted in (c) and (d), respectively. Both $V_P$ and $V_{ISHE}$ peaks appear at the same magnetic field, confirming that the voltages were generated when YIG magnetic moments were preccessing. The sign change in voltage peaks observed between the positive and negative magnetic field regions is consistent with the spin-pumping/ISHE picture.

FIG.5: Microwave power dependent measurements. (a) ISHE voltage measurements with different insertion powers ($P_{MW}$). (b) A plot of ISHE voltage peak to peak amplitude ($V^*_{ISHE}$) as a function of $P_{MW}$. $V_{ISHE}$ scales with $P_{MW}$ as expected from the spin pumping theory in the linear regime.

FIG.6: In-plane (a) and out-of-plane (b) angular dependences of $V_{ISHE}$ signal peak to peak amplitude ($V^*_{ISHE}$). Fit curves in both angular dependences are discussed in the main text. We show fit curves with four different spin-relaxation time ($\tau_s$) in (b) to illustrate how the model curve changes with $\tau_s$. The best fit curve was produced with $\tau_s = 11 \pm 2$ ps. We define three angles ($\phi$, $\phi_M$, $\theta$) as depicted in Figure's insets.

FIG.7: $V_{ISHE} t_{NbN}/\rho_{NbN}$ as a function of NbN thickness. We plot $V_{ISHE} t_{NbN}/\rho_{NbN}$ to normalise $V_{ISHE}$ with NbN thickness and resistivity. By using Eq. (3) in the main text, we extract the spin-Hall angle ($\theta_{SH}$) and spin-diffusion length ($l_{sd}$) of NbN to be $1.1 \times 10^{-2}$ and 14 nm. The best fit curve is shown along with the experimental data.

FIG.8: Comparison of (a) ac and (b) dc $V_P$ measurements. The extracted parameters using Equations in Appendix A for each measurement method are depicted in the legends of the figures. We can confirm that the extracted values are almost the same for both measurements.

FIG.9: Eddy-current damping contribution. (a) A schematic of our model for the eddy-current damping. A chain of Magnetic moments (red arrow) lines up along the $y$ direction and we consider the magnetic field at Point $P(x, 0)$. (b) A plot of calculated eddy-current damping as a function of the NbN thickness. The unit of the eddy-current damping is arbitrary in order to discuss them qualitatively. The thickness dependence is clearly different from our experimental results in Fig. 3(b), suggesting that this damping mechanism is not significant in our experiments.

FIG.10: FMR absorption spectra and ISHE voltages as a function of temperature for $t_{NbN}=10$ nm sample. (a) FMR absorption spectra measured at 3 GHz, with temperature ranging from 260K to 3K. (b) Linewidth evolution with temperature for the 3 GHz measurements. Black data corresponds to an YIG/NbN sample and red data to a bare YIG sample. (c) ISHE voltages measured at 3 GHz for the temperature region of 50K-300K. We confirm that the peak height is below the signal-to-noise ratio around 50 K. (d) The normalised ISHE voltage amplitude as a function of temperature. The inset represents our four point probe measurements of NbN resistivity (for $t_{NbN} = 10$ nm).

Figure 1:

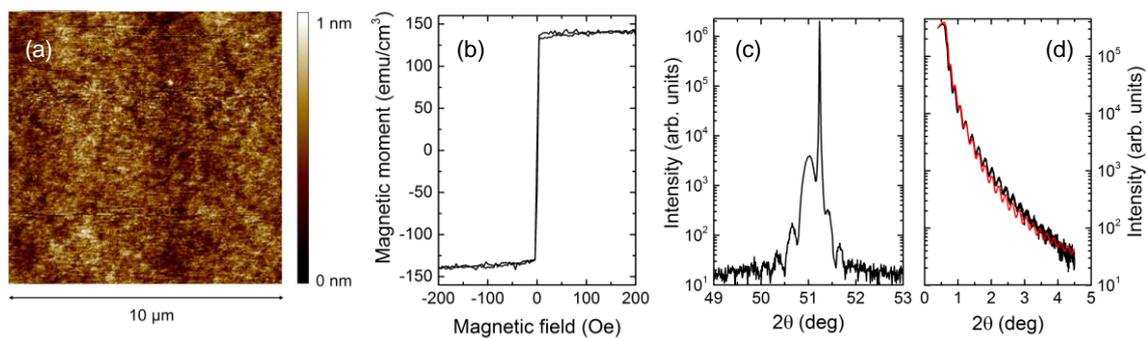

Figure 2:

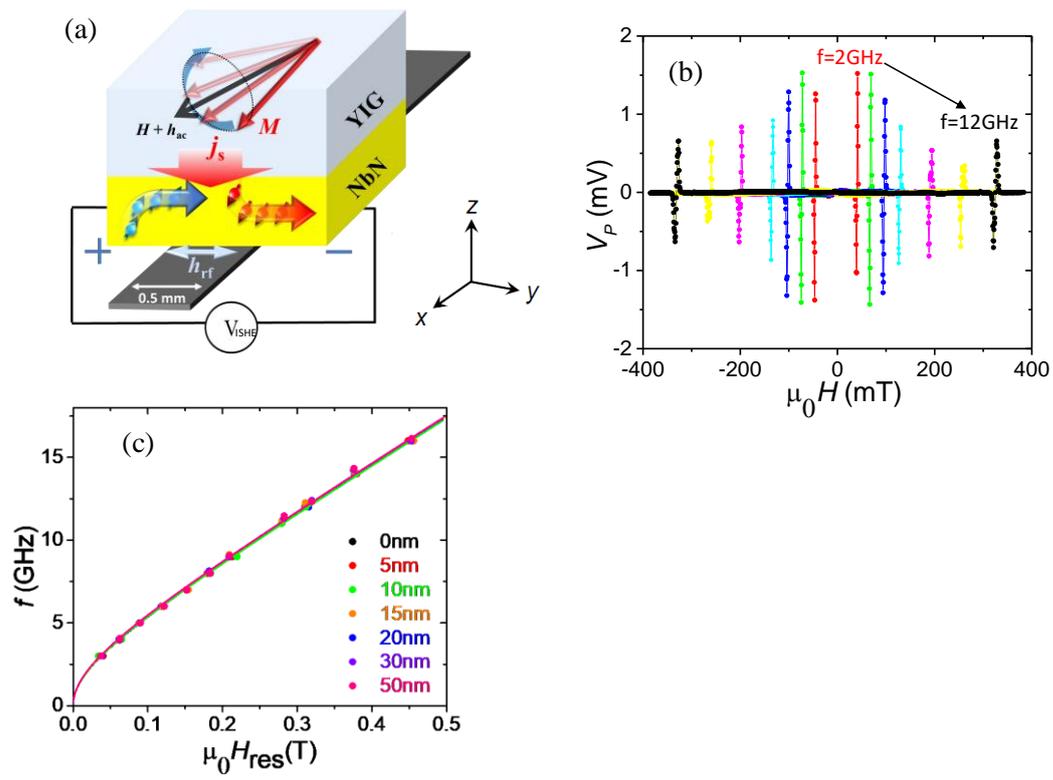

Figure 3:

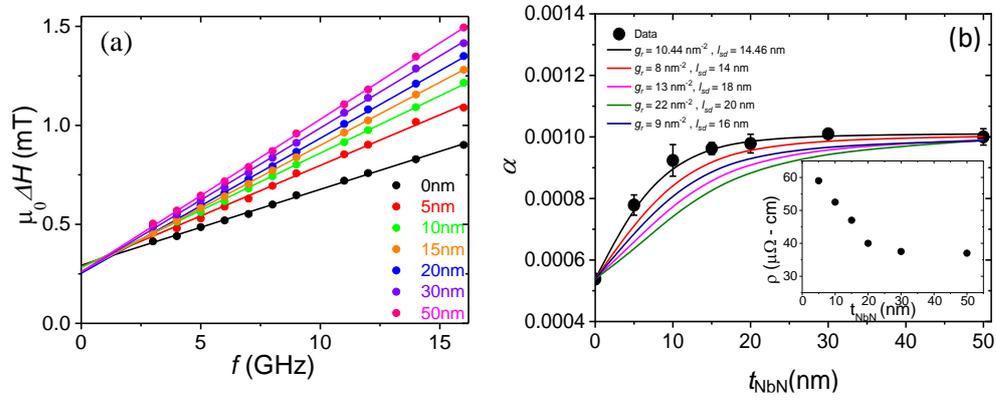

Figure 4:

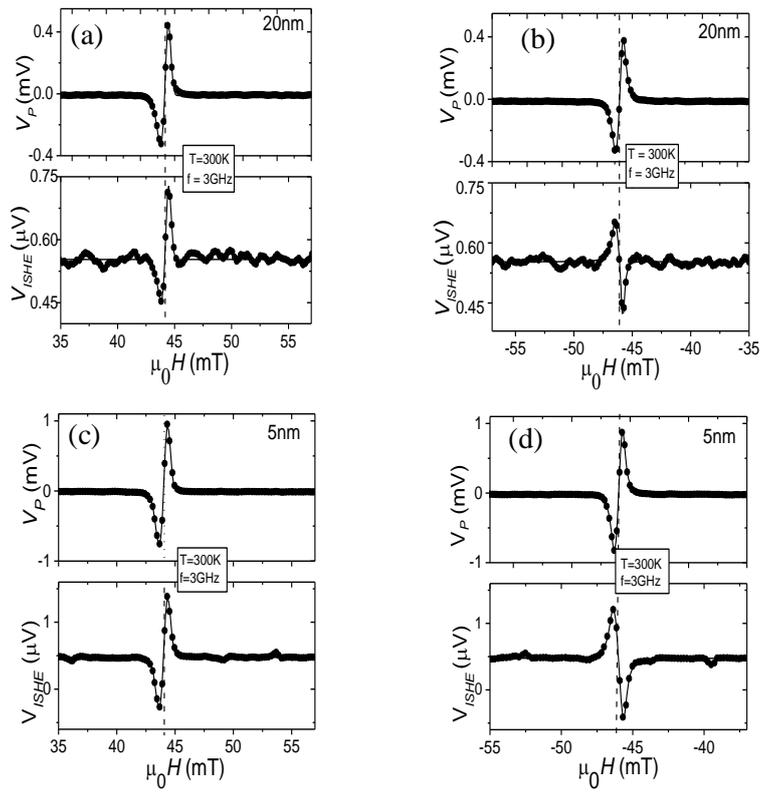

Figure 5:

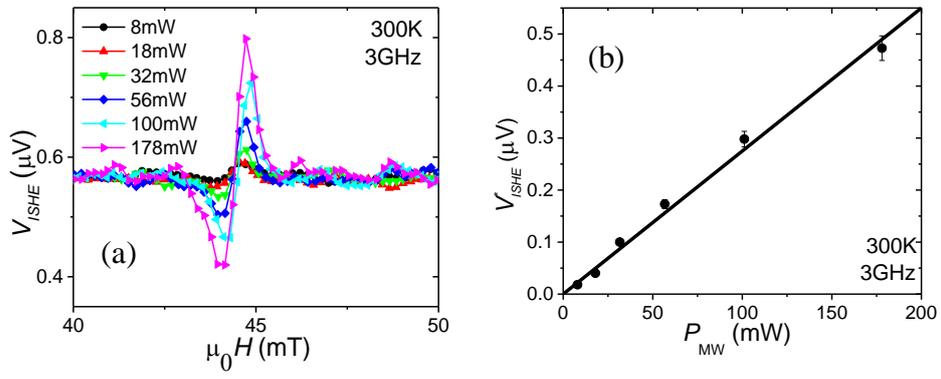

Figure 6:

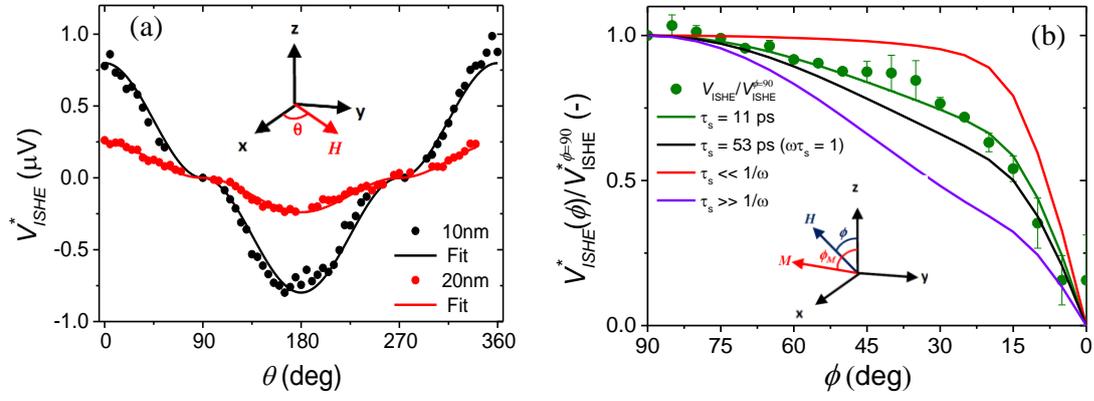

Figure 7:

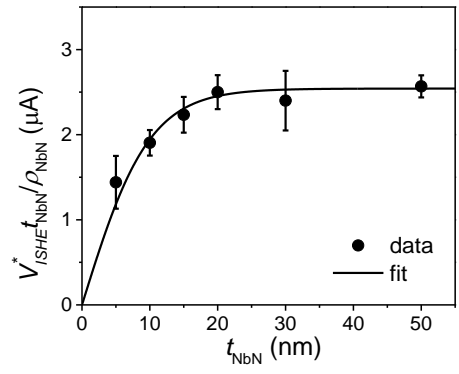

Figure 8

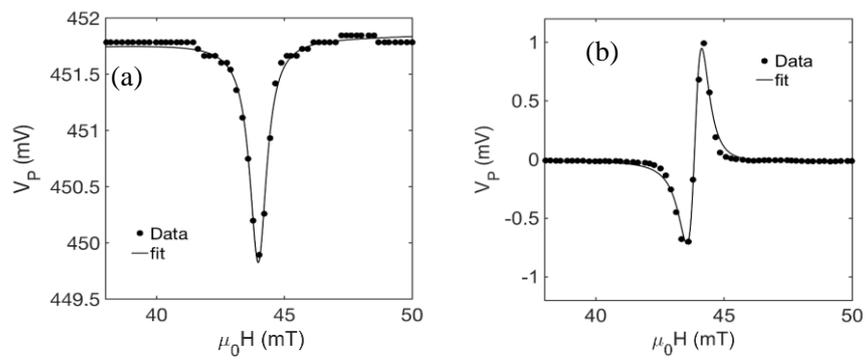

Figure 9:

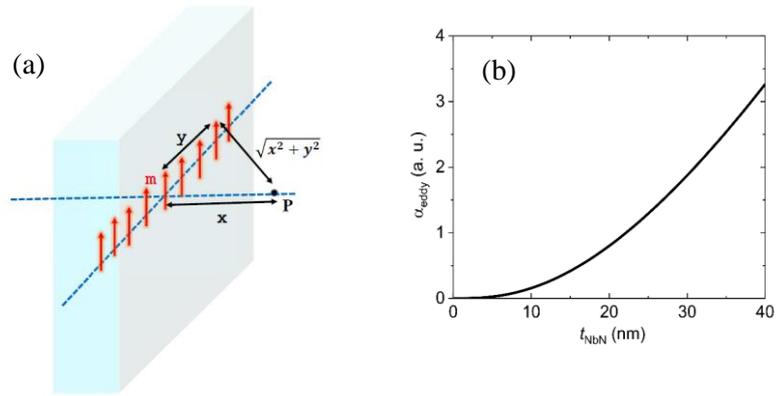

Figure 10:

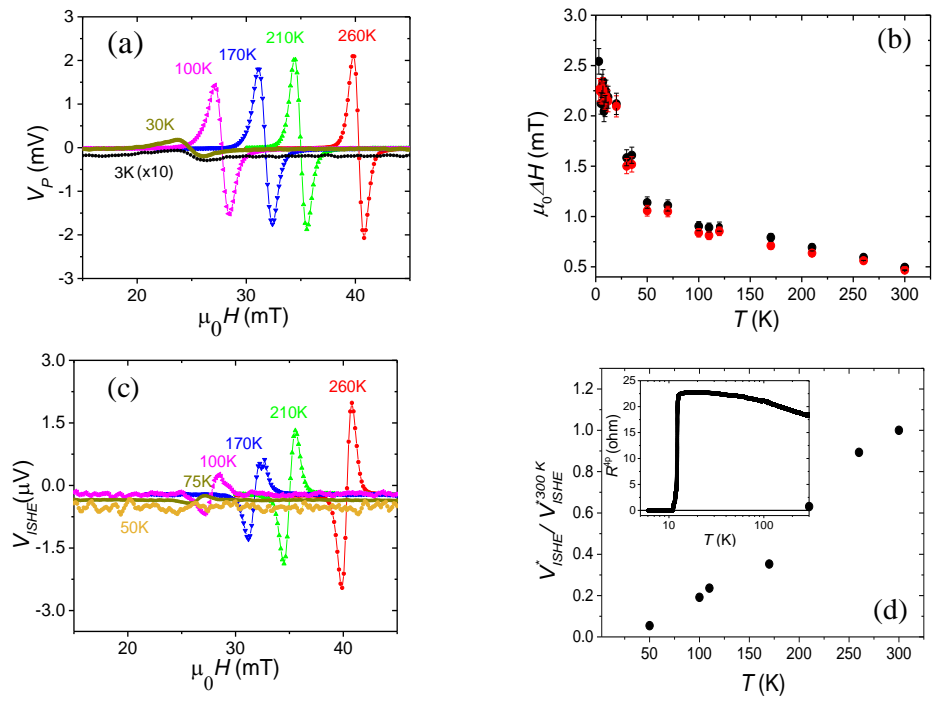